\documentstyle[preprint,12pt,aps,prabib]{revtex}

\begin{document}

\tightenlines
\preprint
{$\ ^{\tt DFPD\ 97/TH/49\ 
(University\ of\ Padova)}_{\tt 
UM-P-97/61\  (University\ of\ Melbourne)}$}

\title{{{\bf 
$p p \leftrightarrow \pi^+ d$ process at low energy:
\\
Interplay between s-- and p--wave mechanisms}}}
\author{
{L. Canton$^{1,2}$, A. Davini$^2$, and P.J. Dortmans$^3$}
\\{\it $^1$Istituto Nazionale di Fisica Nucleare, Sezione di Padova, Italy }
\\ {\it $^2$Dipartimento di Fisica, Universit\`a di Padova, Italy}
\\ {\it $^3$School of Physics, University of Melbourne, Australia}}

\maketitle

\begin{abstract} 

{The large variety of experimental data
around the pion--production threshold are compared  
with a meson--exchange isobar model which includes 
the pion--nucleon interaction in s-- and p--waves.
Theoretical results obtained with two different NN
potentials (Bonn and Paris) indicate that
the behavior of the excitation function 
at threshold is sensitive to
the details of the NN correlations. 
The complete model presented, while developed
originally to reproduce the reaction around
the $\Delta$ resonance, is shown to describe well
the integral (Coulomb--corrected) cross--section
at threshold along with its angular distribution.
At low energies the angular dependence of the analyzing power $A_{y0}$
is well reproduced also. Finally, the energy
dependence of the analyzing power for $\theta=90^o$ 
from threshold up to the $\Delta$ resonance is considered
and discussed. }
\end{abstract}


\newpage
\section{Introduction} 

Pion production in nucleon--nucleon (NN) collisions
at energies near threshold
have attracted a large amount of interest
in recent years.
This interest was triggered by considerable advances 
in experimental techniques~\cite{may90-92,bon96}
which gave a $pp\rightarrow \pi^o pp$
cross--section surprisingly larger than
what was predicted by the established
threshold theory~\cite{kolrei66} of $\pi$N s--wave interaction.
In order to explain this discrepancy, a mechanism employing the 
short--range components of phenomenological NN 
potentials~\cite{leeris93} was introduced to give
sufficient enhancement in the cross--section at threshold
in terms of NN contributions
to the axial charge operator. This effect 
has been recast in terms of explicit heavy--meson exchanges and 
virtual ${\rm N\bar N}$ pair formation in irreducible NN 
production diagrams in the framework of the 
one--boson--exchange theory~\cite{horo94}. 
In both cases, short--range NN correlations have been advocated.
However the same effect has been explained also 
by resorting solely to the properties of the $\pi$N correlations,
and in particular to the off--shell structure of the 
$\pi$N isoscalar amplitude~\cite{heros95}. 
These off--shell extrapolations 
enter in the pion (s--wave) rescattering diagram, and 
as a consequence,
the link between the magnitude of the threshold production cross section
and the $\pi$N scattering lengths
is less direct and cogent than what expected from 
earlier calculations following the on--shell formalism of 
Ref.~\cite{kolrei66}.
A second, independent calculation~\cite{hanh95} analyzed critically
some commonly used approximations and employed
a realistic meson--exchange model for the $\pi$N
T--matrix, with significant differences in the off--shell 
extrapolations.
However, the effect proposed in Ref.~\cite{heros95} was confirmed
but reduced in size, indicating that the correct explanation,
most likely, lies in between the two (NN and $\pi$N) 
effects~\cite{haid96}.

The debate on the missing strength in the $pp\rightarrow\pi^o pp$
cross--section at threshold soon inflamed contiguous reactions,
and in particular the $pp\rightarrow\pi^+d$ 
one where most of the data had been accumulated.
In this case, the threshold rescattering mechanism 
includes charge--exchange and is dominated by the
much larger isovector component of the $\pi$N s--wave amplitude. 
The corrections to this leading
mechanism due to NN (heavy--meson exchange) and $\pi$N
(off--shell) correlations, became a main issue of debate.
First, a great emphasis was put on the
role of heavy--meson exchange diagrams,
since half of the strength has been ascribed to these 
processes~\cite{horo93}. A critical re-analysis reduced the effects of 
heavy--meson NN correlations~\cite{nisk96}, and found that 
s--wave multiple--step mechanisms with intermediate isobar
excitation have an important role even at threshold.
The full inclusion of all these effects actually 
gave an {\em overestimation} in $pp\rightarrow \pi^+ d$. 
In addition, it was observed~\cite{leenucl} 
that the heavy--meson exchange currents
are not so large for  
the $pp\rightarrow \pi^+np$ and $pp\rightarrow \pi^+d$ 
reactions. The smallness of the heavy--meson exchange mechanisms
in this latter channel has been independently confirmed~\cite{haid96}.
A significant increase (50\%) 
in the cross--section was found by inclusion of the 
off--shell structure of the isoscalar $\pi$N amplitude~\cite{hanh97}, 
while the isobar effects in near threshold were not considered.

All these studies deal solely with s--wave pion production mechanisms, 
however p--wave mechanisms must come into play, at a certain stage. 
Such mechanisms
have been advocated for the deviations from the data seen
in the $pp\rightarrow \pi^o pp$ reaction around $\eta\simeq 0.4$
while for the $pp\rightarrow \pi^+ d$ reaction deviations already occur 
around $\eta\simeq 0.2$ ~\cite{haid96} ($\eta$ is the c.m. momentum
of the pion, in units of pion masses). On the other hand,
it has been observed previously~\cite{nisk96} 
that major changes in the 
importance of s--wave mechanism at threshold may have dramatic consequences
not only at low energies but also nearby the $\Delta$ resonance peak
if one looks at the polarization observables, {\em e.g.}
$A_{y0}$, where the interplay between s-- and 
p--wave mechanisms provide the main structure for the observable. 
Global analyses from threshold up to the isobar
resonance have a greater value, but are also much more difficult.

The aim of this paper is to study 
the properties of the $pp\rightarrow \pi^+ d$
reaction in the energy region where
the p--wave mechanisms become relevant, 
and show that it is possible to reproduce the bulk results 
(including spin measurements) for the reaction
from threshold up to the $\Delta$ resonance with a simple
model including s-- and p--wave mechanisms. 
As has been established, the irreducible
heavy--meson diagrams have a small effect in this particular
channel and therefore we do not include these diagrams. 
In the present analysis,
pion production in s--wave is based principally on the 
isovector--dominated
rescattering mechanism triggered by
the $\pi$N--$\pi$N  $\rho$--exchange diagram, while
the p--wave mechanism is dominated by the established
$\Delta$--rescattering diagram. Only the standard corrections
from the $\pi$NN vertex interaction (in both p-- and s--waves)
have been considered here.

We cannot insist upon  the simultaneous reproduction
at threshold of both $\pi$N scattering data 
and $\pi$--production data from NN collisions.
The debate on this point
will eventually be settled amongst $\pi$N off--shell correlations,
role of explicit $\Delta$ degrees in two--baryon s--wave mechanisms,
and perhaps smaller contributions from irreducible
heavy--meson exchange currents. 
Our calculations do not include such effects.
We must note, however, that even in this (simplified) model,
a large sensitivity was found with respect to the nucleonic potential 
employed. In other words, care must be exercised 
with respect to the detailed 
treatment of the {\em conventional} NN correlations calculated within
a DWIA--type framework. Occasionally this sensitivity
has been acknowledged~\cite{heros95,horo93}, in other cases
it has been questioned~\cite{haid96}, but whether it 
masks (partially or totally) any signal of finer effects
in the threshold cross section should be clarified once and for all.
Another aspect of concern should be the sensitivity
with respect to the cut--off of the $\pi$NN vertex in the 
s--wave rescattering diagram. 
There is a general acknowledgment that
the cross section 
is sensitive to the value of this cut--off,
especially the 
$pp\rightarrow \pi d$ cross section.
Choices range from a soft cut--off 
(say, below 800--900 MeV~\cite{haid96,nisk96,leenucl,can96}), 
to a hard cut--off (above 1500--1600 MeV~\cite{hanh95,horo93,dor97}), 
and something in between (1250 MeV~\cite{heros95}).
With respect to the reported cut--off values,
one should add that in Refs.~\cite{horo93,dor97} 
the isovector amplitude was generated explicitly 
by $\rho$--mediated mechanisms, and this allowed 
the use of harder $\pi$NN cut--offs. In particular, 
in Ref.~\cite{horo93} it was set to infinity.
This sensitivity adds a further difficulty
to the disentanglement of any small s--wave correction
in the $pp\rightarrow \pi^+ d$ case.

The present work originated in the 
necessity to provide a tested model
for pion production/absorption which includes the 
sole p--wave and s--wave mechanisms and is 
sufficiently simple but phenomenologically constrained 
for the extension to few--nucleon systems in 
the energy range from pion threshold up to region 
around the isobar--resonance.
It has been shown already that the basic p--wave mechanism
(with explicit allowance of the $\Delta$ resonance),
when tested at the level of the two--nucleon collisions,
can be successfully employed for the description
of the reaction $pd\rightarrow\pi^+ t$ around the 
isobar resonance~\cite{can97}.
However, a study of the spin observables
at this energy~\cite{can-II} indicates that
smaller components from other mechanisms play an important role.
Moreover, with the sole p--wave mechanisms
calculated in Ref.~\cite{can97}, the $pd\rightarrow\pi^+ t$ 
cross section decreases too rapidly with respect to
the data in moving from the $\Delta$ resonance towards
threshold. This is similar to what occurs 
in the $pp\rightarrow\pi^+d$ case.
Obviously, s--wave $\pi$--production mechanisms play an important role 
also in $pd$ collisions, and therefore it is of great importance 
to consider the simultaneous effect of both components.

\section{The Theoretical Model} 
We have calculated the following expression for the 
production/absorption amplitude
\begin{equation}
A = \langle NN^{(-)}|{\cal A}|\pi d \rangle \, ,
\end{equation}
where $|\pi d\rangle$ and $\langle NN^{(-)}|$ describe the
pion--deuteron and NN channel states.
The pion--deuteron state is assumed as free (i.e. asymptotic)
while the NN state represents a two--body scattering 
wave with incoming boundary conditions.  Proper antisymmetrization
with respect to the nucleonic coordinates has been taken into account.
The absorption mechanisms considered  in the calculation
are specified by the detailed structure of the interaction operator
${\cal A}$ and are schematically illustrated in
Fig.\ref{diagrams}. 

The diagram on top of Fig.\ref{diagrams} represents the 
$\Delta$--rescattering mechanism.
It has been calculated starting from the non--relativistic
$\pi$N$\Delta$ interaction Hamiltonian density
\begin{eqnarray}
{\cal H}_{\pi N\Delta}({\bf r})= 
{f_{\pi N\Delta}\over m_\pi}
(\vec{S} \cdot \vec{\nabla}_\pi)
(\vec{\Phi}_\pi({\bf r})\cdot \vec{T})\, .
\label{piND}
\end{eqnarray}
Another necessary ingredient for the determination
of this mechanism is the $\Delta$N--NN transition interaction,
which has been obtained ~\cite{mac89} from the $\pi$-- and 
$\rho$--exchange diagrams
\begin{equation}
V_{N\Delta}=(V_{N\Delta}^\pi+V_{N\Delta}^\rho)
(\vec{T}^+_1\cdot
\vec{ \tau}_2)\, ,
\label{vd}
\end{equation}
with
\begin{equation}
V_{N\Delta}^\pi=-{f_{\pi NN}f_{\pi N\Delta}
\over m_\pi^2} (\vec{S}^+_1\cdot \vec{Q})
(\vec{\sigma}_2\cdot \vec{Q})
\left[{1\over 2\omega_\pi^2}+{1\over
2\omega_\pi^2+2m_\pi(M_\Delta-M)}
\right] \,\, ,
\label{vdpi}
\end{equation}
\begin{eqnarray}
V_{N\Delta}^\rho=-{f_{\rho NN} f_{\rho N\Delta}
\over m_\rho^2} (\vec{S}^+_1\times \vec{Q})\cdot
(\vec{\sigma}_2\times \vec{Q})&\left[{1\over
2\omega_\rho^2}
+ {1\over 2\omega_\rho^2+2m_\rho(M_\Delta-M)}\right]
\nonumber\\
+ {f_{\rho NN} f_{\rho N\Delta} \over (1+\chi) m_\rho^2}
[4i\vec{S}^+_1\cdot (\vec{Q}\times\vec{P}) 
-{(\vec{\sigma}_1\times  \vec{Q}) \cdot (\vec{S}^+_1\times\vec{Q})}]
&\left[{1\over
2\omega_\rho^2}
+ {1\over 
2\omega_\rho^2+2m_\rho(M_\Delta-M)}
\right]\,\, .
\label{vdrho}
\end{eqnarray}

In these expressions, 
$\vec{Q}$ represents the baryon--baryon transferred momentum,
$\vec\sigma$ and $\vec\tau$ denote
the Pauli matrices for the nucleonic spin and isospin, while
$\vec{ S}$ and $\vec{ T}$ are the corresponding
generalization to the nucleon--isobar transition.
In Eq.(\ref{piND})
the baryonic density has been omitted for brevity,
while the pionic isovector field
is denoted by $\vec{\Phi}_\pi({\bf r})$.
The nucleon, pion, and $\rho$ masses
are indicated with $M$, $m_\pi$, and $m_\rho$ respectively,
while $\omega_\pi$ and $\omega_\rho$ represent
the relativistic energy of the two mesons.
These contributions include spin--orbit and other
relativistic corrections to the transition 
potential~\cite{can96,mac89}. 
At each meson--baryon coupling,
form--factors of monopole type are introduced
$(\Lambda^2-m^2)/(\Lambda^2+{Q}^2)$
with the exception of the $\rho N\Delta$ coupling, 
where a dipole--type form factor is assumed.
In the $\Delta N$ exchange diagrams
we have taken into account 
the $\Delta N$ mass difference 
in an approximated way (by considering
the form $2\omega^2+2m(M_\Delta-M)$
instead of the exact
$2\omega(M_\Delta-M+\omega)$
term ) since in this case analytical expressions in partial waves could be 
obtained. Relevant expressions
in partial waves have been given elsewhere~\cite{can96,dor97}
and are not reproduced here.
Finally, the $\Delta$--rescattering mechanism 
requires the specification on how the 
isobar resonance propagates in the intermediate states.
For this purpose, the isobar mass has been endowed 
with an imaginary component linked to the 
resonance width. The detailed structure of the imaginary 
term herein employed has been derived from the study of the 
$\pi^+ d\rightarrow pp$ process around the $\Delta$ 
resonance~\cite{can96}.

The second mechanism depicted in 
Fig.\ref{diagrams} is triggered by the $\pi$NN vertex
and is sometimes referred to as the Impulse Approximation (IA) 
mechanism.
This contribution is calculated starting from the 
non--relativistic pion--nucleon interaction Hamiltonian density
\begin{eqnarray}
{\cal H}_{\pi NN}= {f_{\pi NN}\over m_\pi}
\left(\vec{\sigma} \cdot \left[ \vec{\nabla}_\pi - 
{\omega_\pi\over M} 
\nabla_N^{\!\!\! \!\!\! \!\!\! \! \matrix{\ ^\leftrightarrow}}
\right] \right)
(\vec{\Phi}_\pi({\bf r})\cdot \vec{\tau}) \, .
\label{piNN}
\end{eqnarray}
This form is obtained when performing 
the non--relativistic limit of the pseudo--vector 
coupling between $\pi$ mesons and nucleons~\cite{eri88}. 
The contribution specified by the operator 
$\nabla_N^{\!\!\! \!\!\! \!\!\! \! \matrix{\ 
^\leftrightarrow}}$ is usually referred to as
the Galilei--invariant recoil term
and acts on the nucleonic coordinates
to the right and left according to
the definition 
$
\nabla_N^{\!\!\! \!\!\! \!\!\! \! \matrix{\ ^\leftrightarrow}}
=
(
\nabla_N^{\!\!\! \!\!\! \!\!\! \! \matrix{\ ^\rightarrow}}-
\nabla_N^{\!\!\! \!\!\! \!\!\! \! \matrix{\ ^\leftarrow}}
) / 2
$.

The mechanism on bottom of Fig.\ref{diagrams} includes the 
additional contributions due to the s--wave $\pi$N interaction,
and represents a $\pi$N rescattering process
specified by the following K--matrix structure
\begin{equation}
K_{\pi N} = -{2\over m_\pi}
\left(\lambda_0 + 
\lambda_\rho g_\rho(q)
(\vec{t}_\pi\cdot\vec{\tau})\right)
\, . 
\label{kmat}
\end{equation}
Such an interaction includes both
isoscalar and isovector components. 
The former is originated by the
phenomenological Hamiltonian density
\begin{eqnarray}
{\cal H}_{\pi\pi NN}^{0}= {4\pi\lambda_0 \over m_\pi}
(\vec{\Phi}_\pi({\bf r}) \cdot \vec{\Phi}_\pi({\bf r})),
\label{isoscalar}
\end{eqnarray}
and represents pion rescattering
without charge exchange.
For the latter, which describes $\pi$N scattering 
with charge exchange, we have adopted the $\rho$--meson exchange model
wherein the interaction is entirely given in terms of the
$\rho$--exchange contribution. In 
this case $\lambda_\rho$ and $g_\rho(q)$
are defined by the relevant parameters 
(coupling constants and cut-offs) 
of the $\rho$--exchange vertices,
\begin{equation}
\lambda_\rho = {f_{\rho\pi\pi}f_{\rho NN}\over 8\pi}
{m_\pi^2\over m_\rho^2}\, ,
\end{equation}
and
\begin{equation}
g_\rho(q)=
{m_\rho^2\over m_\rho^2+q^2}
\left(
{\Lambda_\rho^2-m_\rho^2
\over
\Lambda_\rho^2+q^2}
\right)^2\, .
\end{equation}
These contributions have been discussed 
elsewhere~\cite{eri88,ham67} and the 
specialized pion--absorption matrix elements 
have been derived in Ref.~\cite{dor97}.

In calculating the production/absorption mechanism
the unitary effects in the $\pi$N 
system have been taken into account 
through the Heitler equation. 
Such effects have been considered in the framework of pion--nucleon 
scattering, {\it e.g.} in Ref.~\cite{coo87}, 
and herein are applied to the production process.
The resulting (on--shell) T--matrix then becomes
\begin{eqnarray}
T_{\pi N}(q) = -{2\over m_\pi}
&& \left\{ 
\left[
{2\over 3}
\left({\lambda_0+\lambda_\rho g_\rho(q) \over 
   1+2i{q\over m_\pi}(\lambda_0+\lambda_\rho g_\rho(q))}\right)
+
{1\over 3}
\left({\lambda_0-2\lambda_\rho g_\rho(q) \over 
   1+2i{q\over m_\pi}(\lambda_0-2\lambda_\rho g_\rho(q))}\right)
\right] + \right. \nonumber\\
&&
\left.
\left[
{1\over 3}
\left({\lambda_0+\lambda_\rho g_\rho(q) \over 
   1+2i{q\over m_\pi}(\lambda_0+\lambda_\rho g_\rho(q))}\right)
-
{1\over 3}
\left({\lambda_0-2\lambda_\rho g_\rho(q) \over 
   1+2i{q\over m_\pi}(\lambda_0-2\lambda_\rho g_\rho(q))}\right)
\right]
(\vec{t}_\pi\cdot\vec{\tau}) \right\}
\, . 
\label{tmat}
\end{eqnarray}
The results of Eqs.~(\ref{tmat}) and (\ref{kmat})
converge in the threshold limit ($q/m_\pi \rightarrow 0$),
but for higher energies the unitary effects must be included. 

Each meson--baryon vertex 
has been endowed with phenomenological form--factors,
since the sources of the meson fields
are composite objects of extended nature.
For the transition potential, Eqs.(\ref{vd}--\ref{vdrho}),
we have adopted the coupled--channel model III 
given in Ref.~\cite{mac89}. For reference,
the corresponding coupling constants are 
reproduced in Tab.\ref{params}, along with 
all the parameters employed in the
calculations shown herein.
These include $\lambda_0$, the isoscalar strength
of the effective four--leg vertex given by 
Eq.(\ref{isoscalar}), and the effective strength of the
$\rho$--exchange diagram, $\lambda_\rho$.

Finally, the procedure required the setting of 
only one parameter in this study, $\Lambda_B$. 
This cut--off value corresponds to a monopole form factor
and governs the extended structure of both the $\pi$NN and 
$\pi$N$\Delta$ vertices when the pion is on its mass shell. 
Such a form factor 
depends on the baryonic coordinates and has been introduced
in the $\pi d\leftrightarrow pp$ 
process~\cite{dor97} following considerations similar to
those observed previously for the $\pi$N system~\cite{sch94}.
With $\Lambda_B\sim$ 0.7 GeV,
the production cross--section
calculated at the resonance peak
describes well the experimentally measured values.

\section{Results.}
We compare now the theoretical results obtained with the 
model discussed in the previous section
with the low energy experimental data for the $pp\rightarrow \pi^+ d$
process.
Since there are slight differences with respect to our previous 
analyses~\cite{can96,dor97}
we recalculate the integral cross section from threshold
up to the $\Delta$ resonance and beyond. The calculated cross sections are
shown in Fig.\ref{wholeX}. The solid line represents
the calculation obtained with the full model, {\em i.e.} including all
mechanisms discussed herein, and using the Bonn B 
potential~\cite{mach87}
for the evaluation of the two--nucleon initial state 
interaction and of the deuteron wavefunction in the outgoing 
channel. The dotted line describes calculations obtained 
with the full model when the Paris potential~\cite{laco80}
is employed to describe the NN interactions in the incoming and outgoing 
channels.
The dashed line has been obtained using
the Bonn B interaction 
with the s--wave T--matrix contributions set to zero.

The differences between the solid and dashed lines 
indicate that although the $\pi$N $\rho$--exchange 
mechanism dominates the total cross section at threshold, 
at the resonance peak the same mechanism causes a suppression,
due to a destructive interference between it and the 
resonant p--wave process. With respect to this point,
we note that the various mechanisms are often specified by 
the {\em pion--nucleon} state, but in general 
this does not necessarily coincide 
with the state of the {\em pion--nucleus} system, 
the two being related by Jacobian--type angular transforms.
In our approach, we duly calculate the transformations
connecting the different coupling schemes. For this reason,
a large number of NN partial waves are coupled together
by each mechanism, and this may lead to interference effects.

Comparison between the solid and dotted curves 
indicates  that the cross section with the Paris
interaction is smaller than the Bonn result. This behavior has been observed 
previously~\cite{can97} for both the $pp\rightarrow \pi^+ d$
reaction and the more intricate $pd\rightarrow \pi^+ t$
process. In the latter case, the effect is more pronounced.
Over the entire energy spectrum shown,
the calculations made with the full model are in good agreement
with the experimental
data~\cite{hei96,hut91,rit93,for96,rit83,rit81,kor91,data1,data2}.
Around the $\Delta$ resonance,
the differences in the normalization of the cross section 
between the Bonn and Paris calculations can be compensated by a 
slight variation in the cut--off parameter $\Lambda_B$.
Therefore, we draw no conclusion
as to whether one NN potential is preferable to another.
The differences between the two calculations only 
serve to show the sensitivity of the results with 
respect to the details of the model interaction. 

In Fig.\ref{lowX} we display the previous
figure again, but on this occasion with an expanded energy scale
at and above the threshold energy.
However the experimental points shown in this figure 
are not exactly those of Fig.~\ref{wholeX},
since in this figure the data have been corrected
for the Coulomb effects. These effects
diminish rapidly in value with increasing energy,
however all calculations exhibited here 
have been performed without taking such effects into account.

Assuming charge independence, we have considered also  
the data for the $np\rightarrow\pi^o d$ reaction
(scaled by a factor of 2)~\cite{hut91}.
These data have been denoted by triangles.
The solid, dotted, and dashed lines are the same as those
displayed in Fig.\ref{wholeX}. The additional 
(dashed--dotted) curve shows the result obtained using the 
Paris interaction when the $\pi$N s--wave T--matrix is suppressed.
As can be seen, it is possible to achieve agreement
with the experimental data by including all 
mechanisms discussed in the text. However, 
from the difference between
the complete calculations performed with Bonn and Paris 
potentials it is apparent that the threshold--expansion parameters
are very sensitive to which NN potential is employed.
Therefore, in converting from one NN interaction to the other,
it is not possible to reproduce the
behavior of the production cross section at threshold
without a corresponding modification of the parameters 
governing the production mechanisms.
If we expand the purely nuclear cross section
as $\sigma(\eta)=\alpha\eta+\beta\eta^3$,
we find that in passing from Bonn to Paris interaction
the parameter $\alpha$ is reduced by 40\% while
$\beta$ increases by 30\% .
This suggests that if one wishes to use
the Paris potential as the basic
interaction a sizable re-tuning of the parameters
reported in Tab.\ref{params} is essential.
In addition to this remarkable
sensitivity of the low--energy cross section
to the details of the NN potential,
we note that the reaction at threshold is 
dominated by the mechanism triggered by the 
$\pi$N s--wave T--matrix. This is well known
and can be seen directly in Fig.~\ref{lowX}
by comparing the dashed and dotted--dashed lines,
wherein the $\pi$N T--matrix has been set to zero. 
Therefore, we conclude that the process at threshold is 
strongly dependent to {\em both} NN and 
$\pi$N correlations. Furthermore, since the $\beta$ 
coefficient of the cross--section expansion
at threshold is dominated by the p--wave mechanisms
(these include the $\Delta$ rescattering)
it means that all the ingredients 
included in the model have some relevance near threshold
and so cannot be ignored.
Indeed, while the effect due to the p--wave mechanisms below
$\eta=0.1$ is practically negligible,
its contribution rapidly becomes significant, so that by
$\eta=0.4$, the p--wave contribution amounts
to roughly 50\% of the total cross section.
The differences between the two curves 
show that for these p--wave mechanisms the 
sensitivity to the nuclear potential
is somewhat smaller, but remains sizeable.
At low $\eta$ and with the $\pi$N T--matrix set to zero, 
the term which is of greatest importance
is the recoil component in Eq.~(\ref{piNN}). However, 
in the corresponding amplitude there is a cancellation 
between the s-- and d--wave deuteron component 
which reduces the overall impact of the recoil effects
in the cross section~\cite{kolrei66,horo93}.

We stress that our aim is not to obtain a best fit 
to the experimental data at threshold.
Had that been the case, reasonable changes in the parameters
of Tab.~\ref{params} would have led to better fits for 
{\em both} Paris and Bonn results. 
Our main intention is to show that this model, originally
constructed to describe the reaction around the $\Delta$ resonance,
gives quite reasonable results
at lower energies without any need for further refinements. 
However, by considering two equally realistic NN interactions we 
are able to assess
that the results are quite sensitive with respect to
the treatment of the NN correlations.

In Fig.~\ref{diffall} 
the angular distribution of the production cross 
section is reported at various values of $\eta$ 
around threshold.
The theoretical calculations have been performed including all mechanisms
presently discussed, and with the Bonn B potential.
The angular dependence is very well
represented by the theory for various values of $\eta$
ranging from 0.634, down to a minimum of 0.062.
The five curves in the uppermost section of Fig.~\ref{diffall} correspond 
to the theoretical results obtained
for the values of $\eta$ referring to 0.634, 0.443, 
0.350, 0.251, and 0.215.
The points have been extracted from the experimental data 
of Refs.~\cite{rit93,rit81}. 
Similarly, in the middle and bottom parts of the figure
we have compared the resulting angular distributions with the 
experimental analysis of Refs.~\cite{for96} and ~\cite{hei96},
respectively.
The five curves in the middle section correspond 
to the values 0.062, 0.090, 0.13, 0.18, and 0.22 for $\eta$,
while on bottom the curves refer to 0.0761, 0.0951, 0.1240,
0.1434, and 0.202.
We observe that in the energy region considered, the angular 
dependence of the differential cross section is a clear signal 
of the presence of p--wave mechanisms. The sole s--wave mechanisms 
lead here to practically flat (isotropic) cross sections.
The results shown in Fig.\ref{diffall} indicate that 
the p--wave mechanisms are correctly proportioned in this model 
over the whole range considered for the $\eta$ parameter.
Near threshold,
s--wave production is coupled to the $^3P_1$ NN state while 
p--wave production occurs mainly in the $^1D_2$ channel. 
The integral cross--section combines the effects of both
processes, and is therefore more difficult to reproduce than its angular 
distribution. From Fig.\ref{diffall} no conclusion can be drawn about
the total cross--section since the theoretical curves
have been normalized to the data, and such data 
refer to the production of charged pions
in an energy range where Coulomb effects become
significant. In our calculations we have not attempted to estimate
the distortion effects due to the Coulomb interaction, and
therefore 
the comparison with the experimental data 
required a re-normalization of our results.
The normalization factors we have employed for each curve
are given in Tab.~\ref{tablefac}. For comparison, 
the estimated Coulomb suppression factors are displayed also.
For the most part, the normalization factors employed within our 
analysis
are comparable to those obtained from the calculation
of Coulomb corrections.

In Fig.~\ref{Ay0_all} the results for the proton analyzing power $A_{y0}$
at $\eta=0.15$ and $0.21$ are shown in the upper and lower panels
respectively. 
The solid lines represent the full--model 
calculation using the Bonn B potential, while the dotted lines
show the corresponding results obtained with the Paris interaction.
The experimental points have been obtained from Ref.~\cite{kor91}. 
The two figures indicate that the behavior of
$A_{y0}$ is well reproduced around the production threshold
with a model which includes realistic interactions and sensible parameters.
At low energies ({\it i.e.} for $\eta$ $<$ 0.4) the $A_{y0}$ obtained
with the Bonn B potential is smaller in magnitude than the corresponding 
value obtained with the Paris interaction. At $\theta$=90$^o$
the differences between the two curves are largest, 
increasing with increasing energy.
Such behavior suggests that the energy dependence of
$A_{y0}$ at 90$^o$ from threshold up to the $\Delta$ resonance
provides an insightful test for the predictions of the model.
Indeed, at this angle and for low values 
of $\eta$ we find the largest sensitivity to the choice
of NN potential. In addition, this establishes a linkage 
between the low--energy predictions for $A_{y0}$, 
which correctly reproduce the data, and the region around the $\Delta$
resonance, where the calculations tend to overestimate the
experimental results~\cite{can96,dor97}.
Such comparison of $A_{y0}$ at 90$^o$ with the experimental data
is shown in Fig.~\ref{AY90}.
To emphasize the threshold region, we have plotted the proton
analyzing power as a function of $ln (\eta)$.
The range of the horizontal axis covers the entire region from threshold
up to the peak of the $\Delta$ resonance.
The solid line refers to the Bonn B calculation and
the dotted one to the Paris results.
The experimental values
have been obtained from Refs.~\cite{hei96,kor91,data2}.
When the values at exactly 90$^o$ were not directly available,
we have displayed the values 
calculated by interpolation of the nearby data points.
For both interactions, the curves have the correct shape and structure,
although there are differences between the two lines.
For comparison, we show with the dashed curve 
the results obtained only with p--wave mechanisms.
In this case, the results are totally different in structure.

In discussing the calculations for $A_{y0}$, one has implicitly assumed
that the Coulomb interaction does not affect dramatically this 
observable. As a first approximation,
the assumption is correct for both angular distribution of the cross 
section and $A_{y0}$ since the Coulomb penetration factors  in
p-- and s--wave are approximately equal. 
Recent studies~\cite{nisk97} 
have gone beyond that by including the Coulomb distortions
in pion--nucleus wave, finding that the deviations at
$ln(\eta)\simeq -1.4 $ are of the order
$|\delta A_{y0}(90)|=0.04$. They rapidly decrease 
as $\eta$ moves away in both directions.

\section{Summary and Conclusions}

In this paper, the threshold behavior 
of the simplest pion production process,
$pp\leftrightarrow\pi^+ d$,
has been studied by means of 
standard theoretical mechanisms (shown in Fig.\ref{diagrams}), 
which were originally developed in order to describe this reaction 
around the $\Delta$ resonance.
Amongst the various features characterizing the theoretical approach, 
it is worth to mention here that we have employed a 
$\rho$--meson exchange model for the isovector 
$\pi$N coupling, that the unitary effects 
in the $\pi$N correlations have been included, 
and that we have considered
the additional off--shell effects in the intermediate baryonic 
coordinates when the vertices are coupled directly
to the external pion. Then, we have set the 
cut--off governing this off--shell structure, $\Lambda_B$, 
in order to reproduce
the magnitude of the cross section at the resonance peak.
All other parameters remain untouched
with respect to a previous analysis~\cite{dor97}.
We have concentrated this study on the pion production
threshold and
compared the results with measured integral and differential cross--section,
as well as with measurements of proton analyzing powers, $A_{y0}$.

Below $\eta= 0.6$, the reproduction of the
the angular distributions by the complete model 
shows that at low energy 
we describe correctly the fraction of 
p--wave mechanism contributing to the process.
The normalization coefficient of each curve, or 
equivalently the integral cross 
section, is more difficult to reproduce 
since both s-- and p--wave mechanisms
are important, at least until $\eta$ decreases below $0.2$. 
Beyond that, only the s--wave mechanism remains significant
for the cross section.
The normalization coefficients 
are consistent with the estimated Coulomb
suppression factors reported by the phenomenological 
analyses, indicating that both components
of the reaction are reasonably described.

However, 
there are some significant disagreements on the Coulomb estimates,
in the literature.
It may well be that the uncertainty in the Coulomb corrections
is one of the possible reasons for the spreading of the
low--energy data points, as shown in Fig.\ref{lowX}. 
So long as one assumes isospin invariance, 
the data extracted from Ref.~\cite{hut91} is in this respect 
the most reliable, since the Coulomb distortions do not apply.
Curiously, the size of the variation of the calculated results 
with respect to the choice between the two interactions
is roughly comparable to the size of the spreading of 
recently collected data, when Coulomb corrections
are applied. Problems connected with past evaluations 
of Coulomb correction have been emphasized recently~\cite{nisk97}.

The model correctly reproduces the low--energy analyzing power. 
This is a stringent test 
since the observable is governed by interference effects
between amplitudes specific to s-- and p--wave $\pi$N 
mechanisms~\cite{nisk96}. Hence
these processes have to be 
described simultaneously for a correct reproduction of $A_{y0}$.
In addition, the results exhibit a significant dependence
on the choice of the NN interaction, which means 
that NN correlations are important also.

As the energy increase towards the $\Delta$ resonance,
the $A_{y0}$ at $\theta = 90^o$ 
is less well described by the model.
We note a systematic tendency
towards over--estimation when the energy approaches 
the resonance peak. 
Although the gross structure of the observable 
is described qualitatively, 
the effects of other diagrams along with the dynamics in 
higher partial waves have to be described 
with greater accuracy at these (higher) 
energies~\cite{nisk78,blank85,bugg88,arndt93}.
In this respect, the inability of 
the standard (non coupled--channel) meson--exchange NN potentials to
fit relevant NN phase--shifts above pion threshold such as 
the $^3F_3$, must be taken into account or compensated in 
some way, as has been observed recently~\cite{sama97}.

\begin{acknowledgments}
PJD acknowledges the INFN and University of Padova for their
support and hospitality in June--July 1997.
LC thanks the School of Physics in the University of Melbourne
for financial support and hospitality in January--February 1996.
\end{acknowledgments}

\vfill
\eject

\setcounter{figure}{0}
\begin{figure}
\caption{}
Schematic diagrams of the mechanisms included within this analysis.
The upper diagram describes the p--wave $\Delta$--rescattering
mechanism; the middle shows the direct $\pi$NN mechanism; and
the lower diagrams describes the inclusion of the
$\pi$N s--wave interaction.
For all mechanisms, the  NN correlations in the initial 
state are described with the oval on the left, while the deuteron 
wave--function in the final state is represented by the semi oval
on the right.
\label{diagrams}
\end{figure}

\begin{figure}
\caption{}
Total cross section for $\pi^+$ $d$ production (in microbarn) 
from $p p$ collisions. The parameter $\eta$ 
corresponds to the pion momentum in c.m. frame divided by the pion mass.
The full and dotted lines represent
the results obtained with the Bonn B and Paris potentials, 
respectively, and include $\pi$N interaction in p-- and s--waves.
The dashed line shows the effects of excluding 
the $\pi$N T--matrix in s--wave, and calculated with the
Bonn interaction. 
The experimental values have been taken from
Refs.~\cite{hei96,hut91,rit93,for96,rit83,rit81,kor91,data1,data2}.
\label{wholeX}
\end{figure}

\begin{figure}
\caption{}
Total cross section for the production process
at and just above threshold. The solid, dotted and dashed curves 
are equivalent to those of Fig.~\ref{wholeX}.
The dotted--dashed line represents
Paris--potential calculations without including
the $\pi$N interaction in s--wave.
The dots represent the experimental values given in the previous figure,
corrected for the Coulomb effects. 
The triangles denote
the experimental data for the $np\rightarrow\pi^o d$
reaction (multiplied by 2)~\cite{hut91}.
\label{lowX}
\end{figure}

\begin{figure}
\caption{}
Differential cross section at low energy
for the $pp \leftarrow \pi^+ d$ reaction. 
The theoretical model (using Bonn B potential)
is compared with the experimental analysis of 
Refs.~\cite{hei96,rit93,rit83,for96}.
For each curve the corresponding value
of $\eta$ is denoted explicitely.
\label{diffall}
\end{figure}

\begin{figure}
\caption{}
Proton analyzing power (for $\eta$ = 0.15 and 0.21)
with all mechanisms included.
The solid (dotted) curve describes the Bonn B (Paris) results.
The experimental values are taken from from Ref.~\cite{kor91}.
\label{Ay0_all}
\end{figure}

\begin{figure}
\caption{}
Proton analyzing power $A_{y0}$ at $\theta$ = 90$^o$
as a function of $\ln(\eta)$.
The two curves (solid and dotted, respectively)
represent the theoretical results obtained with the Bonn B 
and Paris potentials,
as denoted previously. The dashed line
represents the results obtained in the Bonn case with 
and the p--wave mechanisms only. The experimental points were
taken from
Refs.~\cite{hei96,kor91,data2}.
\label{AY90}
\end{figure}

\newpage

\begin{table}
\begin{center}
\begin{tabular}{|cccc|}
\hline
 & Coupling & Cut--off (GeV) & Formfactor \\
\hline
${\pi N      N}$&${f^2\over 4\pi}$ = 0.0789     & 1.6 & monopole \cr
${\pi N \Delta}$&${f^2\over 4\pi}$ = 0.35       & 0.9 & monopole \cr
${\rho N N }$&${f^2\over 4\pi}$ = 7.61          & 1.2 & monopole \cr
${\rho N \Delta}$&${f^2\over 4\pi}$ = 20.45      & 1.3 &  dipole  \cr
\hline
$\pi \pi N N  $&${\lambda_0}     $ = 0.005      & & \cr 
$\rho$--exch   &${\lambda_\rho}  $ = 0.077      & & \cr 
\end{tabular}
\end{center}
\caption{Parameters used in the calculation. The upper sector
gives couplings and cut--offs for the $\Delta$N--NN 
transition potential (for the $\rho$--meson fields the 
tensor/vector ratio is $\chi$ = 6.1).
The middle sector denotes the parameters
for the effective isoscalar and $\rho$--mediated
$\pi$N interaction is s--wave.
}
\label{params}
\end{table}

\begin{table}
\begin{center}
\begin{tabular}{|cccc|}
\hline
$\eta$ & $E_p^{lab}$ & Normalization & Coulomb factor\\
\hline
0.215 & 294.8 & 0.76 & 0.90 \\
0.251 & 297.5 & 0.92 & 0.91 \\
0.350 & 306.8 & 1.06 & 0.94 \\
0.443 & 317.9 & 1.14 & 0.95 \\
0.634 & 374.4 & 1.07 &  --              \\
\hline
0.062 & 288.1 & 0.60 & 0.74 \\
0.09 & 288.8 & 0.64  & 0.79 \\
0.13 & 290.2 & 0.73  & 0.85 \\
0.18 & 292.7 & 0.66  & 0.89 \\
0.22 & 295.2 & 0.88  & 0.91 \\
\hline
0.0761 & 288.4 & 0.81 & 0.79 \\
0.0951 & 288.9 & 0.85 & 0.84 \\
0.1240 & 289.9 & 0.86 & 0.88 \\
0.1434 & 290.7 & 0.93 & 0.91 \\
0.2023 & 294.1 & 1.02 & 0.94 \\
\hline
\end{tabular}
\end{center}
\caption[*]{Normalization factors for the curves
representing the calculated differential cross section.
The last column shows the estimated Coulomb suppression factor
given in Refs.~\cite{rit93,rit83} (upper section), ~\cite{for96}
(middle section), and ~\cite{hei96} (lower section).}
\label{tablefac}
\end{table}

\end{document}